\documentstyle[a4p,12pt,cite,psfig,mathptm,times]{article}

\newcommand{\WW}{{\rm W}^ + {\rm W}^-}
\newcommand{\ee}{{\rm e}}
\newcommand{\epem}{\ee^+\ee^-}
\newcommand{\zz}{{\rm Z}^0}
\newcommand{\GeV}{\mbox{~GeV}}
\newcommand{\fm}{\mbox{~fm}}
\newcommand{\lll}{\bigg/}
\newcommand{\mcppp}{M^{\pm \pm \pm}(Q_3)}
\newcommand{\mcppn}{M^{\pm \pm \mp}(Q_3)}
\newcommand{\nppp}{n^{\pm \pm \pm}}
\newcommand{\nppn}{n^{\pm \pm \mp}}
\newcommand{\nbppp}{N^{\pm \pm \pm}(Q_3)}
\newcommand{\nbppn}{N^{\pm \pm \mp}(Q_3)}
\newcommand{\delppp}{\delta^{\pm \pm \pm}_2(Q_3)}
\newcommand{\delppn}{\delta^{\pm \pm \mp}(Q_3)}
\newcommand{\delpppt}{\delta^{\pm \pm \pm}(Q_3)}
\newcommand{\delpppg}{\delta^{\pm \pm \pm}_{\rm genuine}(Q_3)}
\newcommand{\pipipi}{\pi^{\pm}\pi^{\pm} \pi^{\pm}}
\newcommand{\pipipin}{\pi^{\pm} \pi^{\pm} \pi^{\mp}}
\newcommand{\rthree}{R_3(Q_3)}
\newcommand{\ronetwo}{R_{1,2}(Q_3)}
\newcommand{\cthree}{C_3(Q_3)}
\newcommand{\ndnb}{\left\langle \frac{\nppp}{\nppn} \right\rangle}
\newcommand{\ndns}{\langle \nppp/\nppn \rangle}
\newcommand{\chidof}{\chi^2/\mbox{d.o.f.}}
\newcommand{\st}{\mbox{~(stat.)}}
\newcommand{\sy}{\mbox{~(syst.)}}

\begin{document}

\begin{titlepage}
\begin{center}
{\large EUROPEAN LABORATORY FOR PARTICLE PHYSICS}
\end{center}
\vspace{2mm}
\begin{flushright}
{\large CERN-EP/98-068 \\ May 7, 1998}
\end{flushright}

\vspace{24mm}

\begin{center}
{\LARGE\bf Bose-Einstein Correlations\\}
\vspace{2mm}
{\LARGE\bf of\\}
\vspace{2mm}
{\LARGE\bf Three Charged Pions in Hadronic Z$^{\bf 0}$ Decays\\}
\end{center}

\vspace{12mm}

\begin{center}
{\LARGE The OPAL Collaboration}
\end{center}

\vspace{24mm}

\begin{abstract}
  Bose-Einstein Correlations (BEC) of three identical charged pions
  were studied in $4 \times 10^6$ hadronic $\zz$ decays recorded with
  the OPAL detector at LEP\@. The genuine three-pion correlations,
  corrected for the Coulomb effect, were separated from the known
  two-pion correlations by a new subtraction procedure. A significant
  genuine three-pion BEC enhancement near threshold was observed
  having an emitter source radius of $r_3 = 0.580 \pm 0.004 \st \pm
  0.029 \sy \fm$ and a strength of $\lambda_3 = 0.504 \pm 0.010 \st\pm
  0.041 \sy$. The Coulomb correction was found to increase the
  $\lambda_3$ value by $\sim 9\%$ and to reduce $r_3$ by $\sim 6\%$.
  The measured $\lambda_3$ corresponds to a value of $0.707 \pm 0.014
  \st \pm 0.078 \sy$ when one takes into account the three-pion sample
  purity. A relation between the two-pion and the three-pion source
  parameters is discussed.
\end{abstract}

\vspace{24mm}

\begin{center}
{\large (Submitted to European Physical Journal, C)}
\end{center}

\end{titlepage}

\begin{center}
{\Large The OPAL Collaboration}
\end{center}

\bigskip

\begin{center}{
K.\thinspace Ackerstaff$^{  8}$,
G.\thinspace Alexander$^{ 23}$,
J.\thinspace Allison$^{ 16}$,
N.\thinspace Altekamp$^{  5}$,
K.J.\thinspace Anderson$^{  9}$,
S.\thinspace Anderson$^{ 12}$,
S.\thinspace Arcelli$^{  2}$,
S.\thinspace Asai$^{ 24}$,
S.F.\thinspace Ashby$^{  1}$,
D.\thinspace Axen$^{ 29}$
% ...
}\end{center}\bigskip
\bigskip
$^{  1}$School of Physics and Astronomy, University of Birmingham,
Birmingham B15 2TT, UK
\newline
$^{  2}$Dipartimento di Fisica dell' Universit\`a di Bologna and INFN,
I-40126 Bologna, Italy
\newline
$^{  3}$Physikalisches Institut, Universit\"at Bonn,
D-53115 Bonn, Germany
\newline
$^{  4}$Department of Physics, University of California,
Riverside CA 92521, USA
\newline
$^{  5}$Cavendish Laboratory, Cambridge CB3 0HE, UK
\newline
$^{  6}$Ottawa-Carleton Institute for Physics,
Department of Physics, Carleton University,
Ottawa, Ontario K1S 5B6, Canada
\newline
$^{  7}$Centre for Research in Particle Physics,
Carleton University, Ottawa, Ontario K1S 5B6, Canada
\newline
$^{  8}$CERN, European Organisation for Particle Physics,
CH-1211 Geneva 23, Switzerland
\newline
$^{  9}$Enrico Fermi Institute and Department of Physics,
University of Chicago, Chicago IL 60637, USA
\newline
$^{ 10}$Fakult\"at f\"ur Physik, Albert Ludwigs Universit\"at,
D-79104 Freiburg, Germany
\newline
$^{ 11}$Physikalisches Institut, Universit\"at
Heidelberg, D-69120 Heidelberg, Germany
\newline
$^{ 12}$Indiana University, Department of Physics,
Swain Hall West 117, Bloomington IN 47405, USA
\newline
$^{ 13}$Queen Mary and Westfield College, University of London,
London E1 4NS, UK
\newline
$^{ 14}$Technische Hochschule Aachen, III Physikalisches Institut,
Sommerfeldstrasse 26-28, D-52056 Aachen, Germany
\newline
$^{ 15}$University College London, London WC1E 6BT, UK
\newline
$^{ 16}$Department of Physics, Schuster Laboratory, The University,
Manchester M13 9PL, UK
\newline
$^{ 17}$Department of Physics, University of Maryland,
College Park, MD 20742, USA
\newline
$^{ 18}$Laboratoire de Physique Nucl\'eaire, Universit\'e de Montr\'eal,
Montr\'eal, Quebec H3C 3J7, Canada
\newline
$^{ 19}$University of Oregon, Department of Physics, Eugene
OR 97403, USA
\newline
$^{ 20}$Rutherford Appleton Laboratory, Chilton,
Didcot, Oxfordshire OX11 0QX, UK
\newline
$^{ 22}$Department of Physics, Technion-Israel Institute of
Technology, Haifa 32000, Israel
\newline
$^{ 23}$Department of Physics and Astronomy, Tel Aviv University,
Tel Aviv 69978, Israel
\newline
$^{ 24}$International Centre for Elementary Particle Physics and
Department of Physics, University of Tokyo, Tokyo 113, and
Kobe University, Kobe 657, Japan
\newline
$^{ 25}$Institute of Physical and Environmental Sciences,
Brunel University, Uxbridge, Middlesex UB8 3PH, UK
\newline
$^{ 26}$Particle Physics Department, Weizmann Institute of Science,
Rehovot 76100, Israel
\newline
$^{ 27}$Universit\"at Hamburg/DESY, II Institut f\"ur Experimental
Physik, Notkestrasse 85, D-22607 Hamburg, Germany
\newline
$^{ 28}$University of Victoria, Department of Physics, P O Box 3055,
Victoria BC V8W 3P6, Canada
\newline
$^{ 29}$University of British Columbia, Department of Physics,
Vancouver BC V6T 1Z1, Canada
\newline
$^{ 30}$University of Alberta,  Department of Physics,
Edmonton AB T6G 2J1, Canada
\newline
$^{ 31}$Duke University, Dept of Physics,
Durham, NC 27708-0305, USA
\newline
$^{ 32}$Research Institute for Particle and Nuclear Physics,
H-1525 Budapest, P O  Box 49, Hungary
\newline
$^{ 33}$Institute of Nuclear Research,
H-4001 Debrecen, P O  Box 51, Hungary
\newline
$^{ 34}$Ludwigs-Maximilians-Universit\"at M\"unchen,
Sektion Physik, Am Coulombwall 1, D-85748 Garching, Germany
\bigskip\newline

\noindent
$^{  a}$ and at TRIUMF, Vancouver, Canada V6T 2A3
\newline
$^{  b}$ and Royal Society University Research Fellow
\newline
$^{  c}$ and Institute of Nuclear Research, Debrecen, Hungary
\newline
$^{  d}$ and Department of Experimental Physics, Lajos Kossuth
University, Debrecen, Hungary
\newline
$^{  e}$ on leave of absence from the University of Freiburg
\newpage

\section{Introduction}

Bose-Einstein Correlations (BEC) between pairs of identical bosons,
mainly the $\pi^{\pm} \pi^{\pm}$ system, have been extensively studied
in a large variety of interactions and over a wide range of energies
\cite{zajc,wolf}. These correlations, which are present when the
bosons are near to one another in phase space, are used to estimate
the size of the emitter of the particles and more recently in
QCD-based models to describe fragmentation and hadronisation in high
energy reactions \cite{geiger,ba}. The two-pion BEC effect has lately
also been discussed \cite{becww} in connection with the measurement of
the W mass in the reaction $\epem \to \WW \to \mbox{hadrons}$ at LEP2.

In systems of more than two identical bosons, BEC are also expected to
be present. These higher-order correlations may affect the
multi-hadron production and are also of interest in intermittency
studies \cite{inter}. Detection of the ``genuine'' multi-boson BEC is
complicated by the fact that they have to be isolated from the
lower-order boson correlations and therefore require large data
samples. In addition, systems of several identical charged bosons,
placed nearby in phase space, are subject to a relatively large
repulsive Coulomb interaction which may suppress the BEC effect. As a
consequence, only relatively few higher order BEC studies of three and
more charged pions have been reported
\cite{tasso,liu,juricic,na23,minibias,na22,delphi}. In those studies,
mainly due to lack of statistics, it was not possible to isolate and
verify the genuine multi-boson BEC from the lower-order ones.
Attempts have been made to infer from the measured BEC of the
multi-boson systems the individual contribution of each of the higher
order correlations by model dependent formulae. A significant genuine
BEC signal of three identical charged pions, $\pipipi$, has been
reported in a hadron-proton interaction experiment \cite{na22} and
more recently in a LEP experiment \cite{delphi} where, however, the
Coulomb effect was neglected.

Here we report on a BEC study of the $\pipipi$ system carried out with
a large sample of approximately $4 \times 10^6$ hadronic $\zz$ decays,
recorded by the OPAL detector at the $\epem$ LEP collider during the
years 1991 to 1995. In this analysis, which takes into account the
Coulomb effect, we have isolated the genuine three-pion BEC and
estimated the size of the emitter. In Section 2 we introduce the
extension of the two-boson BEC to the system of three identical
bosons. Section 3 is devoted to the procedure used for the Coulomb
correction of the three-pion BEC and in Section 4 the experimental
details are given. In Section 5 we describe our method for the
extraction of the genuine $\pipipi$ BEC and present the results
obtained from our analysis. The relations between the two-pion and the
genuine three-pion BEC parameters are explored in Section 6. Finally,
the summary and conclusions are presented in Section 7.

\section{The three-boson correlation function}

In describing the three-boson BEC we follow the approach which was also
adopted, for example, in \cite{wolf}. The BEC of pairs of
identical bosons can be formally expressed in terms of the normalised
function
\begin{equation}
\label{eq_1}
R_2 \ = \ \frac{\rho_2(p_1,p_2)}{\rho_1(p_1)\rho_1(p_2)} \ = \ \sigma
\frac{d^2\sigma}{dp_1dp_2}\lll\left\{\frac{d\sigma}{dp_1}
\frac{d\sigma}{dp_2}\right\} \ ,
\end{equation}
where $\sigma$ is the total boson production cross section,
$\rho_1(p_i)$ and $d\sigma/dp_i$ are the single-boson density in
momentum space and the inclusive cross section, respectively.
Similarly $\rho_2(p_1,p_2)$ and $d^2\sigma/dp_1dp_2$ are respectively
the density of the two-boson system and its inclusive cross section.
The product of the independent one-particle densities
$\rho_1(p_1)\rho_1(p_2)$ is referred to as the reference density
distribution, or reference sample, to which the measured correlations
are compared. The inclusive two-boson density $\rho_2(p_1,p_2)$ can be
written as:
\begin{equation}
\label{eq_rho2}
\rho_2(p_1,p_2) \ = \ \rho_1(p_1)\rho_1(p_2) + K_2(p_1,p_2) \ ,
\end{equation}
where $K_2(p_1,p_2)$ represents the two-body correlations. In the simple
case of two identical bosons the normalised density function $R_2$,
defined in Eq.~\ref{eq_1}, already describes the genuine two-body
correlations and has been referred to in previous BEC studies of OPAL
\cite{prev,newm} as the $C_2$ correlation function. Thus one has
\begin{equation}
\label{eq_r2}
C_2 \equiv R_2 \ = \ 1 + \stackrel{\sim}{K}_2(p_1,p_2) \ ,
\end{equation}
where $\stackrel{\sim}{K}_2(p_1,p_2) =
K_2(p_1,p_2)/[\rho_1(p_1)\rho_1(p_2)]$ is the normalised two-body
correlation term. Since Bose-Einstein correlation are present when the
bosons are close to one another in phase space, one natural choice is
to study them as a function of the variable $Q_2$ defined by
\[ Q_2^2 = q_{1,2}^2 = -(p_1 - p_2)^2 = M^2_2 - 4\mu^2 \ ,\] 
which approaches zero as the identical bosons move closer in phase
space. Here $p_i$ is the four-momentum vector of the $i$th particle,
$\mu$ is the boson mass and $M^2_2$ is the invariant mass squared of
the two-boson system.

In the parametrisation proposed by Goldhaber et al.\ \cite{goldhaber},
$C_2$ has the form
\begin{equation}
C_2(Q_2) = 1 + \lambda_2 e^{-Q^2_2 r^2_2} \ ,
\end{equation}
where $r_2$ estimates the size of the two-boson emitter which is taken
to be of Gaussian shape. The strength of the BEC effect, frequently
referred to as the chaoticity parameter, is measured by $\lambda_2$
which varies in the range $0 \le \lambda_2 \le 1$.

The inclusive density of three bosons, $\rho_3(p_1,p_2,p_3)$, includes
the three independent boson momentum spectra, the two-particle
correlations $K_2$ and the genuine three-particle correlations $K_3$,
namely:
\begin{equation}
\label{eq_rho3}
\rho_3(p_1,p_2,p_3) \ = \ \rho_1(p_1)\rho_1(p_2)\rho_1(p_3) +
\sum_{(3)} \rho_1(p_i)K_2(p_j,p_k) + K_3(p_1,p_2,p_3) \ ,
\end{equation}
where the summation is taken over all the three possible permutations.
The normalised inclusive three-body density, is then given by
\begin{equation}
\label{eq_r3}
R_3 \ = \ \frac{\rho_3(p_1,p_2,p_3)}{\rho_1(p_1)\rho_1(p_2)\rho_1(p_3)} \ = \
1 + R_{1,2} + \stackrel{\sim}{K}_3(p_1,p_2,p_3) \ .
\end{equation}
Here
\begin{equation}
R_{1,2} \ = \ \frac{\sum_{(3)} \rho_1(p_i)K_2(p_j,p_k)}
{\rho_1(p_1)\rho_1(p_2)\rho_1(p_3)} \ ,
\end{equation}
represents a mixed three-boson system in which only two of them are
correlated, and
\begin{equation}
\stackrel{\sim}{K}_3(p_1,p_2,p_3) =
\frac{K_3(p_1,p_2,p_3)}{\rho_1(p_1)\rho_1(p_2)\rho_1(p_3)} \ ,
\end{equation}
represents the three-boson correlation. In analogy with $C_2$, one can
define a correlation function $C_3$, which measures the genuine
three-boson correlation, by subtracting from $R_3$ the term which
contains the two-boson correlations contribution. Thus
\begin{equation}
\label{eq_c3}
C_3 \ \equiv \ R_3 - R_{1,2} \ = \ 1 + \stackrel{\sim}{K}_3(p_1,p_2,p_3) \ ,
\end{equation}
which depends only on the genuine three-boson correlations. For the
study of the three-boson correlations we use the variable $Q_3$ which
is defined as
\[ Q^2_3 \ = \ \sum_{(3)}q^2_{i,j} \ = \ M^2_3 - 9\mu^2 \ , \]
where the summation is taken over all the three different $i,j$ boson
pairs and $M^2_3$ is the invariant mass squared of the three-boson
system. From the definition of this three-boson variable it is clear
that as $Q_3$ approaches zero so do the three $q_{i,j}$ values
which eventually reach the region where the two-boson BEC enhancement
is observed.

The genuine three-pion correlation function $\cthree$ can be
parametrised by the expression \cite{juricic}
\begin{equation}
\label{eq_c3q}
\cthree \ = \ 1 + 2 \lambda_3 e^{-Q^2_3 r^2_3} \ ,
\end{equation}
where $\lambda_3$, which can vary within the limits $0 \le \lambda_3
\le 1$, measures the strength of the three-boson BEC effect and $r_3$
estimates the size of the three-boson emitter. The factor two which
multiplies $\lambda_3$ arises from the presence of two possible
diagrams with exchange of all the identical pions within a triplet. To
extract from the data values for the strength $\lambda_3$ and the
emitter size $r_3$, we modified Eq.~\ref{eq_c3q} to read
\begin{equation}
\label{eq_c3qd}
\cthree \ = \ \kappa (1 + 2\lambda_3 e^{-Q^2_3r^2_3})
(1 + \varepsilon Q_3) \ ,
\end{equation}
where $\kappa$ is a normalisation factor and the linear term $(1 +
\varepsilon Q_3)$ accounts for the long range correlations arising
from charge and energy conservation and phase space constraints.
Higher-order $Q_3$ terms for the long range correlations were found
not to be needed in the present analysis.

\section{The Coulomb correction}

The observed BEC of identical charged bosons is suppressed by the
Coulomb repulsive force. To account for this effect a correction to
the measured BEC distribution is required. If $C_2(Q_2)$ is the
two-boson correlation in the presence of the Coulomb effect then it is
related to the true $C_2^{\rm true}(Q_2)$ through the function
$G_2(Q_2)$, so that
\begin{equation}
C_2(Q_2) = G_2(Q_2) C_2^{\rm true}(Q_2) \ .
\end{equation}
In the case that the reference sample is a Monte Carlo generated data
without the Coulomb effect, $G_2(Q_2)$ can be expressed by the Gamow
factor \cite{gyulassy}:
\begin{equation}
\label{eq_g2}
G_2(Q_2) \ = \ 2\pi \eta/(e^{2\pi \eta} - 1) \ ,
\end{equation}
where $\eta \ = \ \alpha_{\rm em} \ee_1 \ee_2 \mu/Q_2$. Here $\ee_1$
and $\ee_2$ are the charges, in positron units, of the two bosons
having a mass of $\mu$, and $\alpha_{\rm em}$ is the fine-structure
constant. Recently alternative Coulomb corrections
\cite{bowler,biyajima} for the two-boson system have been proposed,
which are based on non-Gaussian parametrisations. These could not be
extended to the three-pion system and therefore were not used in our
analysis.

For a given three charged bosons system, with boson pairs having $Q_2$
values of $q_{1,2}$, $q_{1,3}$ and $q_{2,3}$, the Coulomb correction
$G_3$ can be approximated, in terms of the $G_2$ function by
\cite{juricic}:
\begin{equation}
\label{eq_g33}
G_3(Q_3) \ = \ {\bf \langle}G_2(q_{1,2})G_2(q_{2,3})G_2(q_{1,3})
{\bf \rangle} \ ,
\end{equation}
where the average is taken over all experimentally accessible values
of $q_{i,j}$ which satisfy the condition $Q_3^2 \ =
\sum_{(3)}q^2_{i,j}$~.

The function $G_3(Q_3)$ can be evaluated through Eq.~\ref{eq_g33} or
by using a slightly more precise formulation proposed in
\cite{cramer}. In our analysis the differences between the results of
these two methods were smaller than the statistical errors so that the
simpler method was sufficiently accurate for our purposes.

\section{Experimental setup and data selection}

\subsection{The OPAL detector}

Details of the OPAL detector and its performance at the LEP $\epem$
collider are given elsewhere \cite{opald}. Here we will describe
briefly only those detector components pertinent to the present
analysis, namely the central tracking chambers.

Besides a silicon microvertex detector, the central tracking chambers
consist of a precision vertex detector, a large jet chamber, and
additional $z$-chambers surrounding the jet chamber. The vertex
detector is a $1~\mbox{m}$ long, two-layer cylindrical drift chamber
that surrounds the beam pipe\footnote{A right-handed coordinate system
  is adopted by OPAL, where the $x$-axis points to the centre of the
  LEP ring, and positive $z$ is along the electron beam direction. The
  angles $\theta$ and $\phi$ are the polar and azimuthal angles
  respectively.}. The jet chamber has a length of $4~\mbox{m}$ and a
diameter of $3.7~\mbox{m}$. It is divided into 24 sectors in $\phi$,
each equipped with 159 sense wires parallel to the beam ensuring a
large number of measured points even for particles emerging from a
secondary vertex. The jet chamber also provides a measurement of the
specific energy loss, $dE/dx$, of charged particles \cite{dedx}. A
resolution of $3-4\%$ on $dE/dx$ has been obtained, allowing particle
identification over a large momentum range. The $z$-chambers,
$4~\mbox{m}$ long, $50~\mbox{cm}$ wide and $59~\mbox{mm}$ thick, allow
a precise measurement of the $z$-coordinate of the charged tracks.
They cover polar angles in the region $|\cos(\theta)| \leq 0.72$ and
$94\%$ of the azimuthal angular range. All the chambers are contained
in a solenoid providing an axial magnetic field of 0.435 T. The
combination of these chambers leads to a momentum resolution of
$\sigma_{p_t}/p_t \approx \sqrt{(0.02)^2 + (0.0015 \cdot p_t)^2}$,
where $p_t$ in $\mbox{GeV}/c$ is the transverse momentum with respect
to the beam direction. The first term under the square root sign
represents the contribution from multiple Coulomb scattering
\cite{jetchamber}.

\subsection{Data selection}

The analysis was performed with the OPAL data collected at LEP with
centre of mass energies on and around the $\zz$ peak with the
requirement that the jet and $z$-chambers were fully operational. The
hadronic $\zz$ decays were selected according to the number of charged
tracks and the visible energy of the event \cite{eventselection}. We
applied the same track quality and $dE/dx$ cuts described in a former
OPAL BEC study of two identical charged pions \cite{prev}.
Furthermore, events with a thrust angle of $|\cos(\theta_{\rm
  thrust})| > 0.82$ with respect to the beam axis were rejected.
Finally we accepted only events with a relative charge balance of
$|(n^+ - n^-)|/(n^+ + n^-) < 0.25$, where $n^+$ and $n^-$ are
respectively the observed numbers of positively and negatively charged
tracks. Following these criteria a total of $2.65 \times 10^6$
hadronic $\zz$ decay events were used in the analysis.

To avoid configurations with overlapping tracks, we rejected pion
pairs if their invariant mass was less than $0.4 \GeV$ and if the
opening angle between them in the plane perpendicular to the beam axis
was less than $0.05~\mbox{rad}$. In the present analysis all charged
tracks are assumed to be pions. From Monte Carlo (MC) studies
\cite{eventselection} we have estimated that the average pion purity
of our charged track sample is $89.3\%$ with a systematic uncertainty
of $\pm 2.2\%$ and a negligible statistical error. Thus the pion
purity of the three charged track system is $71.3 \pm 5.3\%$.

\section{Analysis and results}

\begin{center}
\begin{figure}[htbp]
\centerline{\psfig{figure=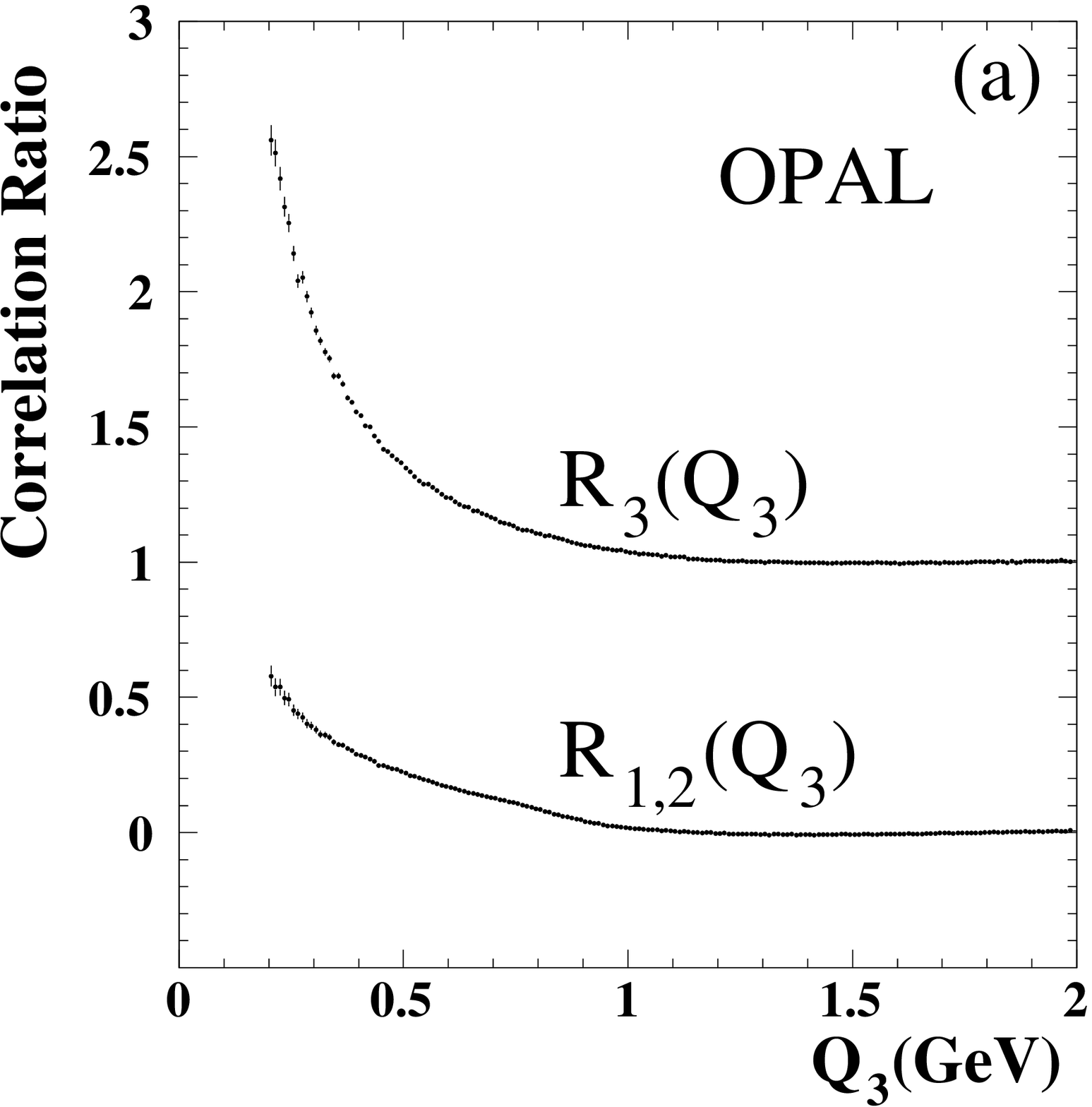,width=10cm,height=10cm}}
\centerline{\psfig{figure=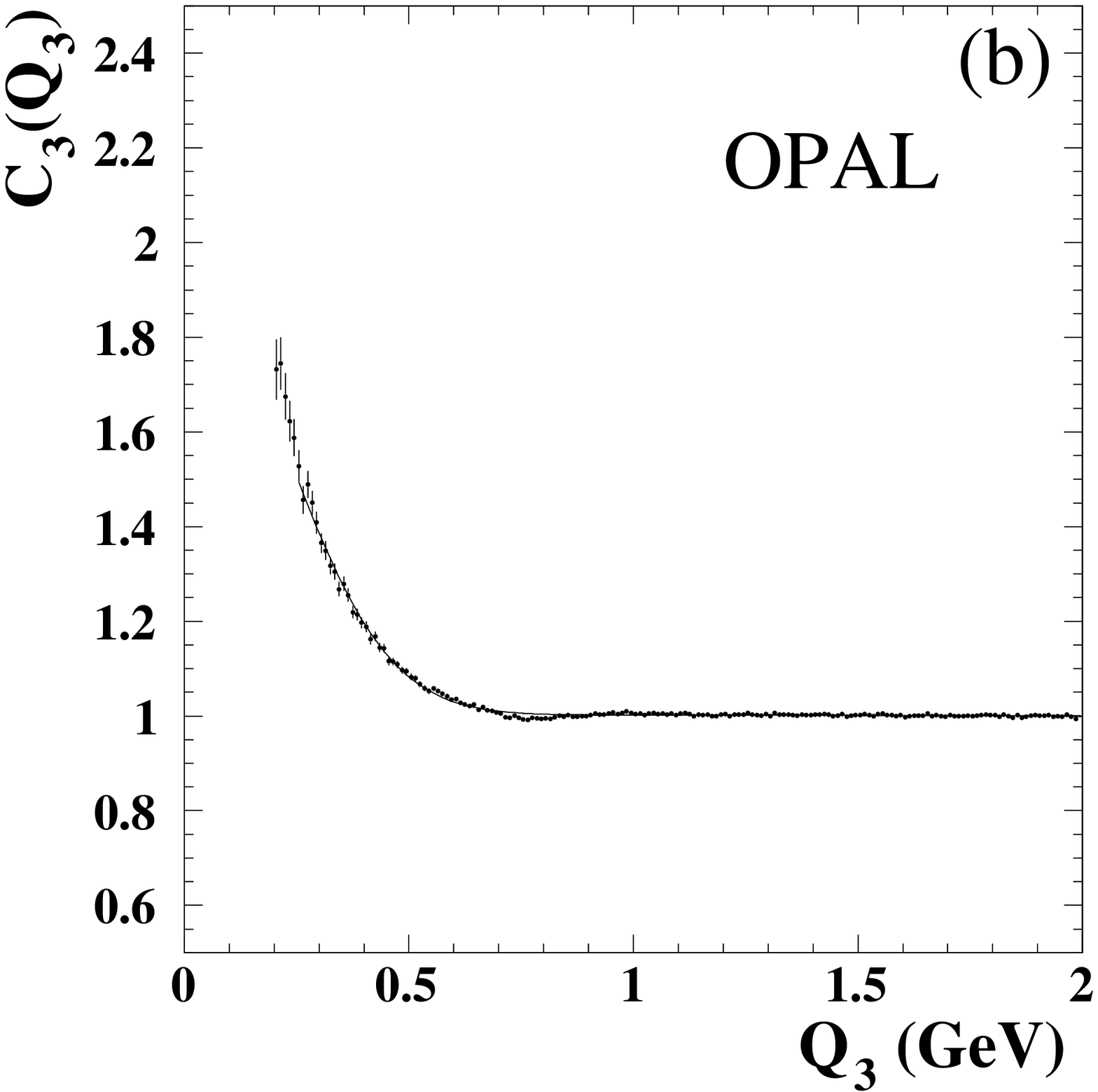,width=10cm,height=10cm}}
\caption{The $\pipipi$ BEC measured distributions,
  without Coulomb correction, as a function of $Q_3$. (a) the measured
  $\rthree$ and $\ronetwo$ distributions before subtraction of the
  two-pion BEC and (b) the $\cthree$ distribution after the
  subtraction of the two-pion BEC. These measurements are represented by
  points with statistical error bars. The solid line in (b) represents
  the fit result of Eq.~\ref{eq_c3qd}, in the range $0.25 < Q_3 < 2.0
  \GeV$, to the measured $\cthree$ distribution.}
\label{fig_3p}
\end{figure}
\end{center}

The BEC analysis used the hadronic $\zz$ decay data of OPAL where in
each event all possible $\pipipi$ combinations were taken as the data
sample. For the reference distribution we used a Monte Carlo generated
sample \cite{jetset} of $4 \times 10^6$ JETSET~7.4 events which have
passed a full detector simulation \cite{allison} but do not include
BEC and Coulomb effects. This JETSET Monte Carlo program, which
includes most of the known resonances which decay into the $\pi^+
\pi^-$ and $\pi^{\pm}\pi^{\pm}\pi^{\mp}$ final states, was carefully
tuned to the OPAL data \cite{para}. Here one should note that the
\mbox{$\pi^{\pm}\pi^{\pm}\pi^{\mp}$} systems of the measured data
contain one pair of identical pions and therefore cannot be used as a
reference sample. Thus:
\begin{equation}
\label{eq_r3d}
\rthree \ = \ \frac{\nbppp}{\mcppp} \ ,
\end{equation}
where $\nbppp$ is the number of the $\pipipi$ data combinations and
$\mcppp$ is the corresponding number of Monte Carlo $\pipipi$
combinations at the same $Q_3$ value. The Monte Carlo sample was
normalised to the data in a $Q_3$ region far away from any observable
BEC enhancement. This was achieved by requiring that the integrated
number of the Monte Carlo entries in the $Q_3$ range $1.6-2.0 \GeV$
was equal to that of the data. The measured $\rthree$ distribution is
shown in Fig.~\ref{fig_3p}a in the range of $0.2 < Q_3 < 2.0 \GeV$.
The data points below $0.25 \GeV$ have relatively large errors, and a
lower three-track separation efficiency of identical charged tracks.
Therefore a lower limit of $0.25 \GeV$ was imposed on the analysis. In
the figure a clear enhancement is observed in the region below $Q_3 =
1 \GeV$. This enhancement can be interpreted as coming from both the
known two-pion and from the genuine three-pion BEC.

\subsection{The extraction of the genuine three-pion BEC}

To extract the genuine three-pion BEC one has to subtract from
$\rthree$ the contribution coming from the well known two-pion BEC. In
our analysis this last contribution is evaluated from the
mixed-charged $\pipipin$ combinations of the data. To this end, it is
convenient to rewrite $\rthree$ as follows:
\begin{equation}
\rthree \ = \ \frac{\nbppp}{\mcppp} \ = \ 1 + \frac{\nbppp - \mcppp}
{\mcppp} \ = \ 1 + \frac{\delpppt}{\mcppp} \ .
\end{equation}
The total excess $\delpppt$ above the Monte Carlo expectation has two
contributions. The first from the two-pion BEC, $\delppp$, and the
second from the genuine three-pion BEC, $\delpppg$, so that:
\begin{equation}
\delpppt \ = \ \delppp + \delpppg \ .
\end{equation}
From this it follows that what should be subtracted from $\rthree$ to
obtain the genuine three-pion BEC, is:
\begin{equation}
\ronetwo \ = \ \frac{\delppp}{\mcppp} \ .
\end{equation}
Because we utilise for the subtraction the same events used for the
$\rthree$ measurement, there exists in every event with given charged
multiplicity $m$ and charge balance $\Delta = |n^+ - n^-|$ values a
relation between $\delppp$ and $\delppn = \nbppn - \mcppn$. Here
$\nbppn$ is the number of $\pipipin$ combinations in the data and
$\mcppn$ is the corresponding number of combinations for the MC
generated sample properly normalised to the data in the $Q_3$ range of
$1.6 - 2.0 \GeV$.

If we define $\nppn$ as the number of $\pipipin$ combinations in a
given event, then it can be related to $\nppp$, the number of $\pipipi$
combinations. A straightforward combinatorial calculation yields that
for an event with given $m$ and $\Delta$ values, one has:
\begin{equation}
\frac{\nppn}{\nppp} \ = \ 3 \left(1 + 4 \frac{m-\Delta^2}
{m^2-4m+3\Delta^2}\right) \ .
\label{eq_bfor}
\end{equation}
In our analysis we utilise, after the application of the appropriate
cuts, all the hadronic $\zz$ decays lying within a wide range of
multiplicity and charge balance. If we define $\ndns$ as the ratio
$\nppp/\nppn$ averaged over all multiplicity and charge balance values
in our data, then:
\begin{equation}
\int \delppp \, dQ_3 \ = \ \ndnb \times \int \delppn \, dQ_3 \ .
\label{eq_delpppi}
\end{equation}
\noindent
The integrations are carried out over the $Q_3$ region where $\delppp$
and $\delppn$ are different from zero. We further verified from MC
studies that, to a good approximation, relation (\ref{eq_delpppi})
holds also in its differential form, that is:
\begin{equation}
\delppp \ = \ \ndnb \times \delppn \ .
\label{eq_delppp}
\end{equation}

\begin{figure}[htb]
\centerline{\psfig{figure=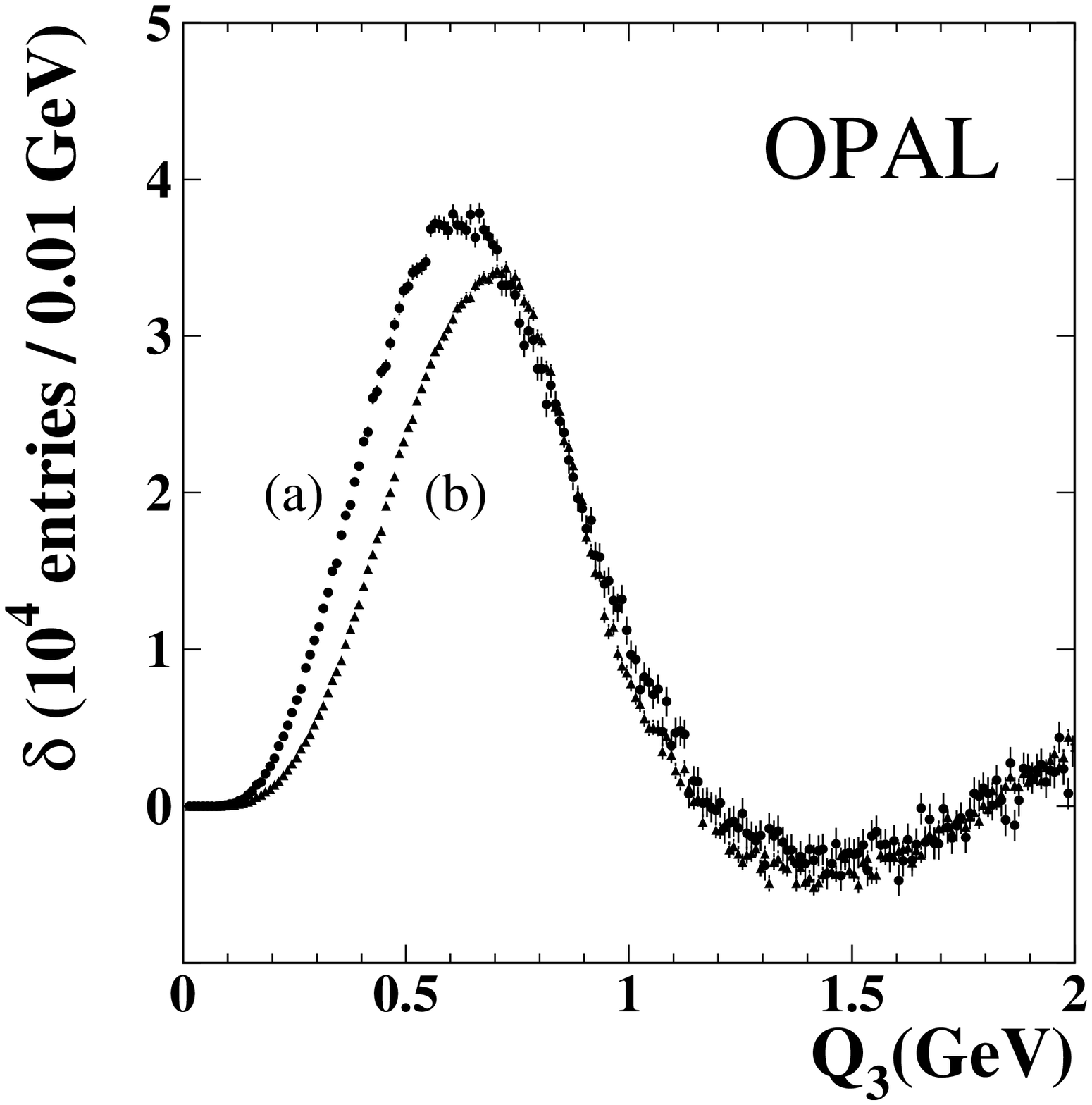,width=12cm,height=12cm}}
\caption{(a) $\delpppt$ and (b) $\ndns \times \delppn$
  as functions of $Q_3$. The errors plotted are the statistical ones.
The difference between the distributions (a) and (b), in the range
$Q_3 < 0.7 \GeV$, is due to the genuine three-pion BEC.}
\label{fig_del}
\end{figure}

The average $\ndns$ depends on the multiplicity $m$ and the charge
balance $\Delta$ distributions of the data sample. Using the MC
hadronic $\zz$ decay sample we determined, by counting the number of
combinations in the $Q_3$ range of $1.6-2.0 \GeV$, that $\langle
\nppn/\nppp \rangle = 3.69$ with a negligible statistical error. The
same value is obtained when the MC sample is replaced by the data
sample. This value shifts to $3.71$ when the $Q_3$ range is enlarged
to $1.5 - 2.0 \GeV$. We also studied the variation of this ratio on
the pion purity. To this end we evaluated this ratio from the MC
sample using only tracks which were generated as pions with the result
that $\langle \nppn/\nppp \rangle = 3.70$. The effect of the shift
from 3.69 to 3.71 on the BEC parameters was found to be negligible in
comparison to the statistical errors and to other systematic
uncertainties (see Table \ref{tab_sys}).

In Fig.~\ref{fig_del} we show the $\delpppt$ and the $\ndns \times
\delppn$ distributions as a function of $Q_3$. As can be seen, the two
distributions are similar in the higher $Q_3$ region, namely between
$0.7$ and $2.0 \GeV$. The slight difference between the two
distributions can be attributed to the systematic uncertainties given
in Table \ref{tab_sys}, in particular those listed as items (f) and
(g). Thus in this $Q_3$ range the excess of $\pipipi$ combinations is
fully accounted for by the excess seen in the $\pipipin$ due to the
two-pion BEC\@. In the lower $Q_3$ region an excess of $\delpppt$ over
$\ndns \times \delppn$ is observed which can no longer be attributed
to the two-pion BEC and is therefore identified as the genuine
three-pion BEC contribution, $\delpppg$. Thus:
\begin{equation}
\cthree \ = \ \rthree - \ronetwo \ = \ \frac{\nbppp}{\mcppp} -
\frac{\delppn}{\mcppp} \times \ndnb \ .
\label{eq_cnew}
\end{equation}
In the present analysis we have used Eq.~\ref{eq_cnew} to subtract the
contributions due to the two-pion BEC\@.

The measured distribution $\ronetwo$ is shown in Fig.~\ref{fig_3p}a
and the resulting $\cthree$ distribution is shown in
Fig.~\ref{fig_3p}b, where a significant genuine three-pion BEC
enhancement is clearly present. The solid line in Fig.~\ref{fig_3p}b
represents the fit result of Eq.~\ref{eq_c3qd} to the data. The fitted
values of $\lambda_3$ and $r_3$ and the correlation factor
$\rho_{\lambda,r}$ are given in Table~\ref{tab_fit} together with the
$\chi^2$ value divided by the number of degrees of freedom (d.o.f.).

\begin{table}[htbp]
\begin{center}
\begin{tabular}{|l||c|c|}
\hline
Parameter & Without Coulomb Corr. & With Coulomb Corr.
\cr
\hline\hline
$\lambda_3$ & $0.462 \pm 0.012$ & $0.504 \pm 0.010$
\cr
$r_3$~[fm] & $0.616 \pm 0.005 $ & $0.580 \pm 0.004 $
\cr
$\kappa$ & $1.003 \pm 0.001$ & $1.026 \pm 0.001$
\cr
$\varepsilon$~[GeV$^{-1}$] & $-0.001 \pm 0.001$ & $-0.015 \pm 0.001$
\cr
\hline
$\rho_{\lambda,r}$ & $+0.887$ & $+0.883$
\cr
\hline
$\chidof$ & $218/171$ & $190/171$
\cr
\hline
\end{tabular}
\caption{Results of the fit of Eq.~\ref{eq_c3qd} to the measured
  $\cthree$ distributions without (Fig.~\ref{fig_3p}b) and with
  (Fig.~\ref{fig_genuine}b) Coulomb correction carried out over the
  range of $0.25 < Q_3 < 2.0 \GeV$. The fitted values are the genuine
  three-pion emitter $r_3$ and the BEC strength $\lambda_3$ together
  with the long range correlation parameter $\varepsilon$ and the
  normalisation factor $\kappa$. The errors are the statistical ones
  obtained by the fits, and $\rho_{\lambda,r}$ is the correlation
  factor between $\lambda_3$ and $r_3$. The quality of the fits are
  presented by their $\chidof$ values.}
\label{tab_fit}
\end{center}
\end{table}

\subsection{Evaluation of the Coulomb effect}

The Coulomb correction to the genuine BEC, defined as $G_3(Q_3)$ in
Eq.~\ref{eq_g33}, can be applied either to the data or to the MC
reference sample. These two possibilities are not expected to yield
identical results since, unlike the MC generated sample, the data are
affected by both the BEC and the Coulomb interactions. In our BEC
analysis we chose to apply the Coulomb correction to the data and
utilised the results coming from the second possibility as a measure
of the systematic errors.

\begin{figure}[htbp]
\centerline{\psfig{figure=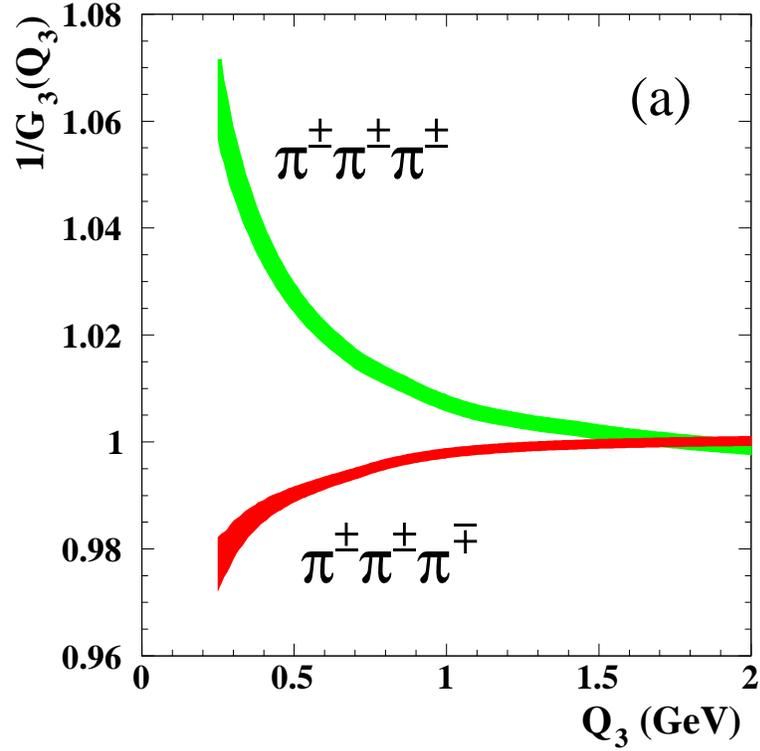,width=10cm,height=10cm}}
\centerline{\psfig{figure=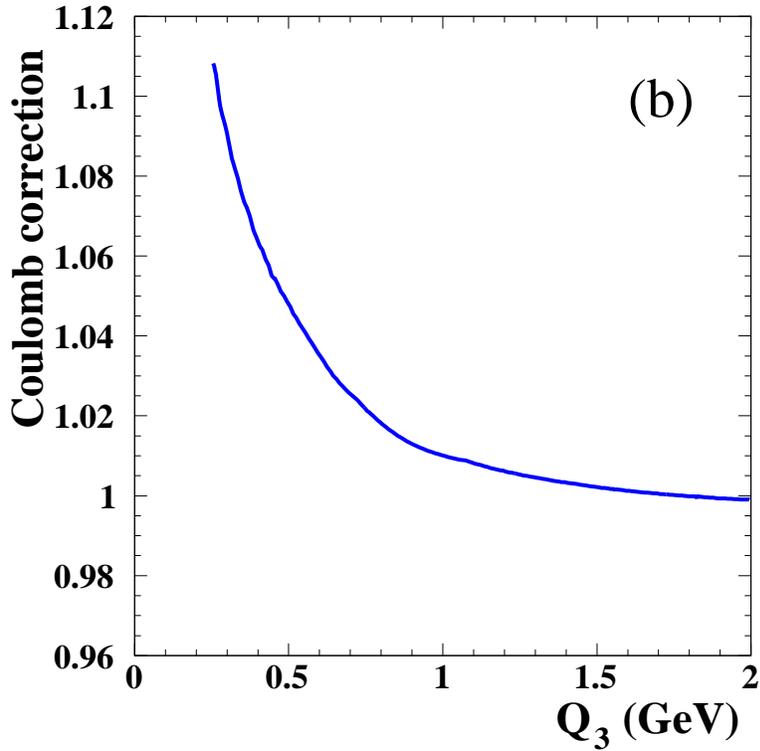,width=10cm,height=10cm}}
\caption{(a) The Coulomb correction factors as a function
  of $Q_3$ for the $\pipipi$ and for the $\pipipin$ distributions. The
  width of the bands corresponds to the statistical uncertainty. (b)
  The mean net Coulomb correction value as a function of $Q_3$ for the
  genuine three-pion BEC distribution obtained by dividing the
  corrected $\cthree$ distribution by the uncorrected one.}
\label{fig_coulomb}
\end{figure}

The Coulomb correction was accounted for by assigning to every
three-pion combination of the data a weight equal to $1/(G_2(q_{1,2})
G_2(q_{2,3}) G_2(q_{1,3}))$. This automatically assures that
$G_3(Q_3)$, defined in Eq.~\ref{eq_g33}, is averaged only over all
accessible experimental values of $q_{i,j}$. Using this procedure, the
Coulomb correction factor $1/G_3(Q_3)$ can be evaluated at every $Q_3$
bin separately for the $\pipipi$ and the $\pipipin$ data samples.
These correction factors are shown in Fig.~\ref{fig_coulomb}a as a
function of $Q_3$ in the range from $0.25$ to $2.0 \GeV$. The
resulting Coulomb correction applied to $\cthree$, due to the
corrections to $\rthree$ and $\ronetwo$, is shown in
Fig.~\ref{fig_coulomb}b where it is seen to rise as $Q_3$ decreases,
reaching the value of about $11\%$ at $Q_3 = 0.25 \GeV$.

\subsection{The corrected $\pipipi$ BEC distributions}

\begin{figure}[htbp]
\centerline{\psfig{figure=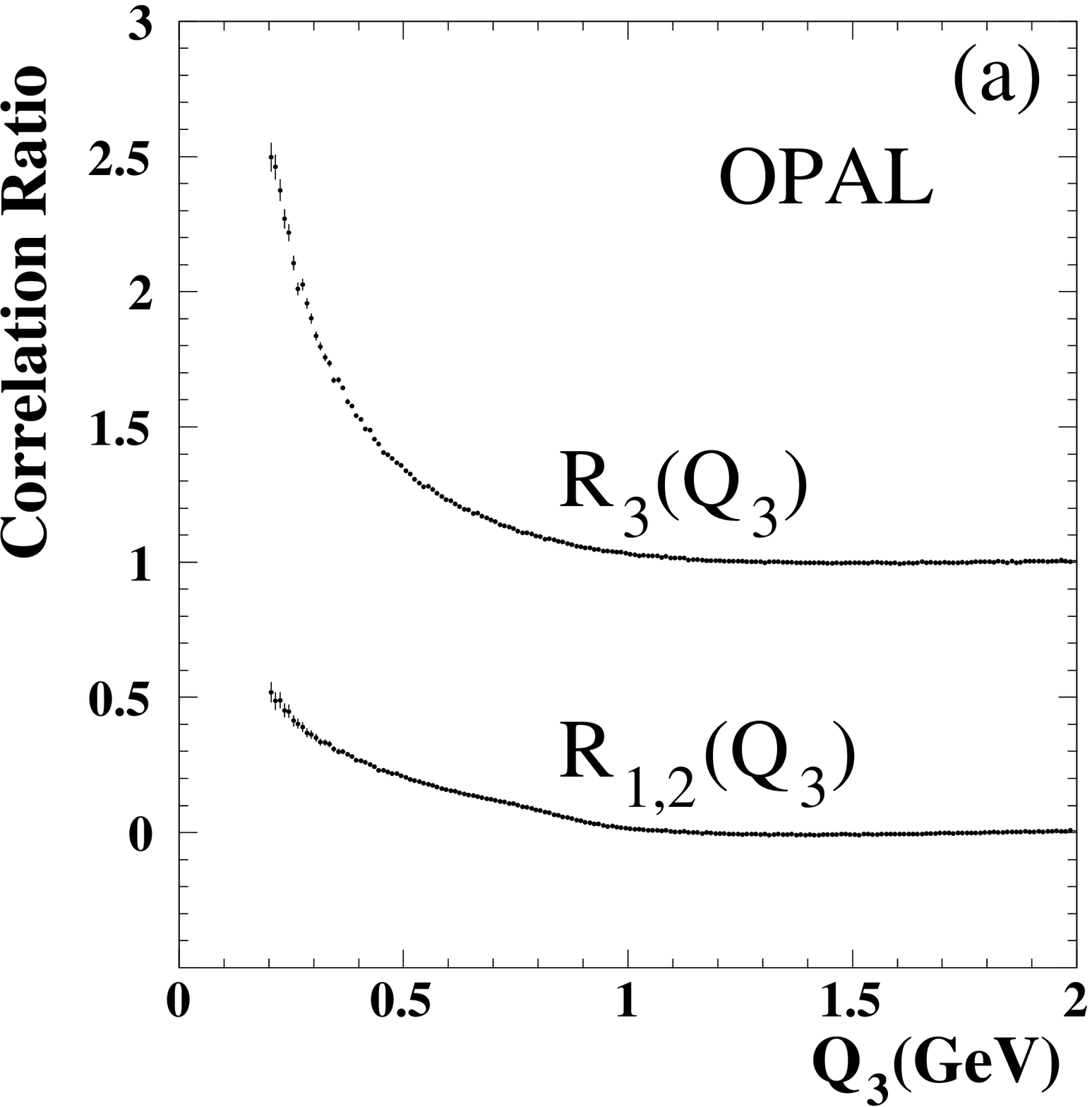,width=10cm,height=10cm}}
\centerline{\psfig{figure=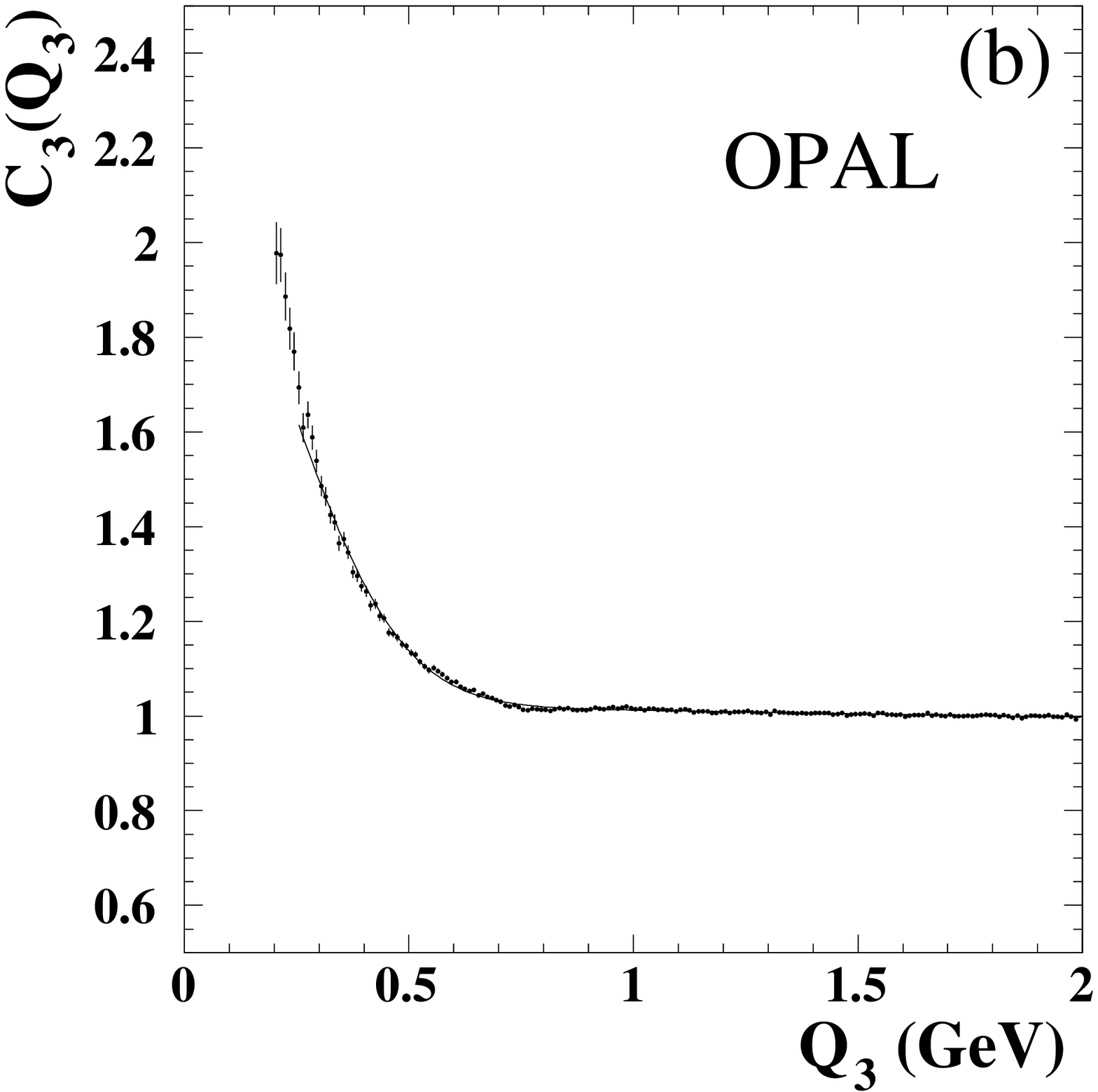,width=10cm,height=10cm}}
\caption{The Coulomb corrected $\pipipi$ BEC correlation distributions.
  (a) The non-subtracted $\rthree$ and $\ronetwo$ distributions, and
  (b) the genuine $\cthree$ distribution. The solid line in (b)
  represents the fit results of Eq.~\ref{eq_c3qd} over the range $0.25
  < Q_3 < 2.0 \GeV$.}
\label{fig_genuine}
\end{figure}

The Coulomb corrected distributions $\rthree$, $\ronetwo$ and
$\cthree$ of the three-pion BEC, are shown in Fig.~\ref{fig_genuine}.
A clear genuine three-pion Bose-Einstein enhancement is present in the
low $Q_3$ region, from about $Q_3 = 0.7 \GeV$ reaching a value of
$\cthree = 2.0$ at $Q_3 = 0.2 \GeV$. The continuous line in the figure
represents the fit result of $\cthree$ in the range $0.25 < Q_3 < 2.0
\GeV$, as parametrised in Eq.~\ref{eq_c3qd}. The quality of this fit,
given by a $\chidof = 190/171$, represents an improvement over the fit
result obtained for the Coulomb uncorrected $\cthree$ distribution.
The values obtained from the fit are $r_3 = 0.580 \pm 0.004 \fm$ for
the emitter radius and a strength of $\lambda_3 = 0.504 \pm 0.010$,
with correlation factor of $\rho_{\lambda,r} = +0.883$. These are
listed in Table~\ref{tab_fit} together with the fit results for
$\kappa$ and $\varepsilon$. A comparison between the results presented
in the table shows, as expected, that the value of $\lambda_3$
increases when the Coulomb correction is applied. We found that
$\lambda_3$ increased by about $9\%$ whereas $r_3$ decreased by about
$6\%$.

\subsection{Systematic errors}

To estimate the systematic errors we have considered the effects on
the $\lambda_3$ and $r_3$ results arising from the choice of the data
selection criteria and from the procedure adopted for the Coulomb
correction. We also investigated the effect on the results from our
choice of the fitting range and the MC reference sample. These are
summarised in Table \ref{tab_sys}.

\begin{table}[htbp]
  \begin{center}
    \begin{tabular}{|l||c|c||c|c|}
      \hline
      Fit variation & $\lambda_3$ & $\Delta \lambda_3$ & $r_3 ~[\mbox{fm}]$ &
      $\Delta r_3~[\mbox{fm}]$
      \cr
      \hline\hline
      a. The reference fit
      & $0.504 \pm 0.010$ & --- &$0.580 \pm 0.004$ & ---
      \cr
      \hline
      b. varying data selection cuts
      & $0.519 \pm 0.017$ & $+0.015$ & $0.561 \pm 0.007$ & $-0.019$
      \cr
      c. MC Coulomb correction
      & $0.488 \pm 0.011$ & $-0.016$ & $0.586 \pm 0.004$ & $+0.006$
      \cr
      d. fit range $0.25 < Q_3 < 1.5 \GeV$
      & $0.507 \pm 0.011$ & $+0.003$ & $0.582 \pm 0.005$ & $+0.002$
      \cr
      e. fit range $0.30 < Q_3 < 2.0 \GeV$
      & $0.471 \pm 0.011$ & $-0.033$ & $0.569 \pm 0.005$ & $-0.011$
      \cr
      f. addition of a long range $Q_3^2$ term
      & $0.506 \pm 0.010$ & $+0.002$ & $0.582 \pm 0.004$ & $+0.002$
      \cr
      g. varying the MC reference sample
      & $0.496 \pm 0.026$ & $-0.008$ & $0.597 \pm 0.015$ & $+0.017$
      \cr
      \hline\hline
      Total systematic error
      & --- & $0.041$ & --- & $0.029$
      \cr
      \hline
    \end{tabular}
\caption{Results of several fits of the $\cthree$
  parametrisation to the data given with their statistical errors.
  The differences, $\Delta \lambda_3$ and $\Delta r_3$, between the
  parameter values of the reference fit (a) and the others from (b) to
  (g) are added in quadrature to obtain an estimate of the combined
  systematic uncertainty associated with the fitted $\lambda_3$ and
  $r_3$ values. These are given in the last row.}
    \label{tab_sys}
  \end{center}
\end{table}

In order to estimate the systematic effects related to track and event
selection the analysis was repeated restricting the track selection
criteria described in Section 4.2. To evaluate the possible
contribution from our choice to account for the Coulomb effect on the
BEC we also applied the Coulomb correction to the MC reference sample
rather than to the data. Furthermore we considered for the fit two
alternative $Q_3$ ranges. We also investigated the influence of adding
a quadratic $Q_3$ term to the long range correlations. Finally, the
systematic uncertainty coming from our choice of the JETSET~7.4 MC
reference sample has been estimated by repeating the analysis with the
HERWIG~5.8 \cite{herwig} generated sample. This alternative MC uses a
totally different model of fragmentation (cluster fragmentation) from
that used by JETSET~7.4 (string formation and fragmentation). An
estimate of the over-all systematic uncertainties was obtained by
summing in quadrature the differences between each fit (b)--(g) and
the reference fit (a).

The largest contributions to the over-all systematic error come from
the choice of the selection criteria and from the choice of the lower
$Q_3$ fit range limit. As seen in Table~\ref{tab_sys}, the results for
$r_3$ and $\lambda_3$ change by less than $4\%$ when the Coulomb
correction is applied to the Monte Carlo generated sample. We observe
that $\lambda_3$ is rather sensitive to the choice of the lower $Q_3$
limit used as compared to the change of $r_3$. In addition to the list
given in Table~\ref{tab_sys}, we also investigated other possible
sources of systematic effects, such as the choice of the $Q_3$ bin
size used in the fit, and verified that they indeed have negligible
contributions. Finally, as noted above, the influence of the
uncertainties of the ratio $\langle \nppn/\nppp \rangle$ on the
subtraction formula and the fit results is also negligible.

\begin{figure}[htb]
\centerline{\psfig{figure=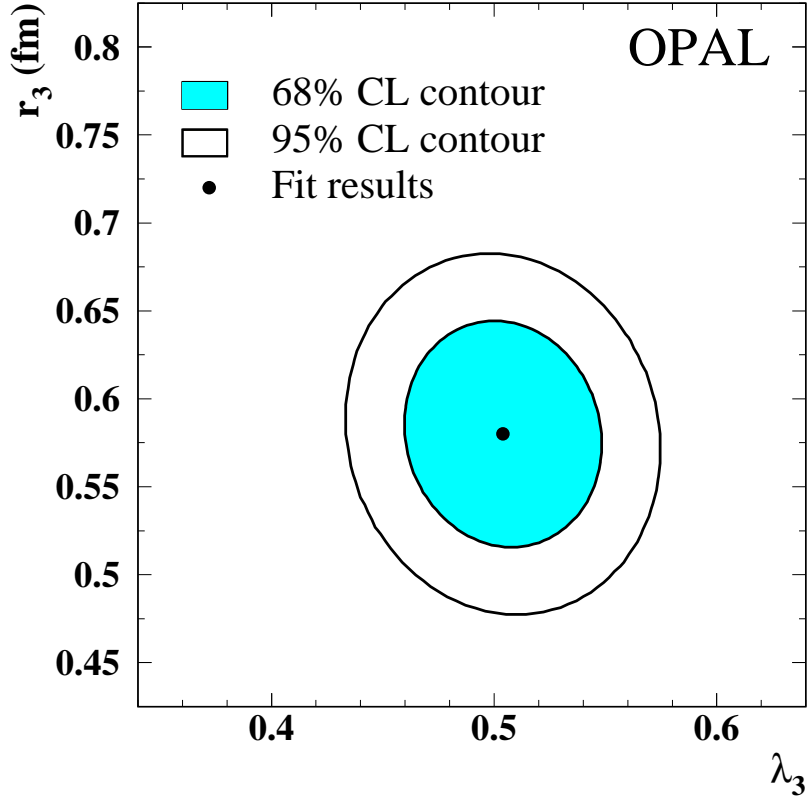,width=12cm,height=12cm}}
\caption{The $68\%$ and $95\%$ confidence level correlation contours
  for the genuine three-pion BEC parameters $\lambda_3$ and $r_3$
  after Coulomb correction. The contours are calculated from the
  statistical errors and the systematic uncertainties listed in Table
  \ref{tab_sys}. The best values are represented by the solid circle.}
\label{fig_cont}
\end{figure}

Thus the final values of the BEC parameters are: $r_3 = 0.580 \pm
0.004 \st \pm 0.029 \sy \fm$ and $\lambda_3 = 0.504 \pm 0.010 \st \pm
0.041 \sy$, where the uncertainties of the measured parameters are
strongly dominated by the systematic errors. In Fig.~\ref{fig_cont}
the $68\%$ and $95\%$ confidence level correlation contours for the
BEC parameters are shown. The shape of the contours is determined
mostly from the systematic errors. Accounting for the three-pion
purity of $0.713 \pm 0.053$, the BEC strength amounts to
$\lambda_3^{\rm pure} = 0.707 \pm 0.014 \st \pm 0.078 \sy$ for a
$100\%$ pure $\pipipi$ system. The purity error of $0.053$, due to the
uncertainty of the MC generation rates, is incorporated in the
systematic error of the $\lambda_3^{\rm pure}$ value.

\section{Relation to other experimental results}

A relation between the two-pion and three-pion emitter radii is
derived in Ref.~\cite{juricic} based on the Fourier transform of the
source distribution which is assumed to be of a Gaussian shape. This
relation confines the emitter range $\stackrel{\sim}{r_3}$ as
determined from a fit to $\rthree$:
\begin{equation}
r_2/\sqrt{3} \ \leq \ \stackrel{\sim}{r_3} \ \leq \ r_2/\sqrt{2} \ ,
\end{equation}
where $r_2$ is the two-boson BEC emitter size. Since in our analysis
the emitter radius is determined from the genuine BEC distribution
$\cthree$, the previous bounds reduce to the equality:
\begin{equation}
\label{eq_r3e}
r_3 \ = \ r_2/\sqrt{2} \ .
\end{equation}
In a former OPAL analysis \cite{newm}, of the two-pion BEC present in
the hadronic $\zz$ decays, two options were adopted for the reference
sample. The first utilised the correlations of the pairs of $\pi^+
\pi^-$ in the data and the second used the two-pion correlations of
Monte Carlo generated sample. Since in our analysis the reference
samples were taken from the Monte Carlo generated events, we have
checked relation (\ref{eq_r3e}) with the previously measured OPAL
value, obtained by the second method, of $r_2 = 0.793 \pm 0.015 \fm$,
where only the statistical error was given. From Eq.~\ref{eq_r3e} it
follows that this $r_2$ value corresponds to $r_3^{\rm calc} = 0.561
\pm 0.011 \fm$, which is in good agreement with our value obtained
from a fit to the data of $r_3 = 0.580 \pm 0.004\st \pm 0.029\sy \fm$.

The BEC of the $\pipipi$ system has also been studied by the DELPHI
collaboration \cite{delphi} using the hadronic $\zz$ decays measured
at LEP\@. In that analysis, which neglected the Coulomb correction,
the following results were obtained: $r_3 = 0.657 \pm 0.039 \st \pm
0.032 \sy \fm$ and $\lambda_3 = 0.28 \pm 0.05 \st \pm 0.07 \sy$. The
relatively large statistical errors reflect the fact that a much
smaller data sample was used than in the present measurement. Our
result for $r_3$, before Coulomb correction, of $0.616 \fm$ with a
statistical error of $\pm 0.005 \fm$, is smaller than the DELPHI
result but is consistent with it within one standard deviation. The
measured $\lambda_3$ value depends on the pion track purity of the
hadron sample analysed, so that caution has to be exercised when
comparing the BEC strength values of different experiments. Keeping
this in mind, we note that our mean $\lambda_3$ value lies
considerably above that reported by DELPHI but it is still consistent
with it within two standard deviations when the systematic errors are
included.

\section{Summary and conclusions}

The Bose-Einstein correlations of three identical charged pions,
produced in hadronic $\zz$ decays, have been studied after correcting
for the Coulomb interaction. A significant genuine three-pion
Bose-Einstein correlation signal is observed near threshold in the
$\cthree$ distribution obtained after the subtraction of the two-pion
correlation contribution. The radius $r_3$ of the three-pion emitter
and the BEC strength $\lambda_3$ are measured to be:
\[ r_3 = 0.580 \pm 0.004 \st \pm 0.029 \sy \fm
\ \ \ \ \ \mbox{and} \ \ \ \ \ \lambda_3 = 0.504 \pm 0.010 \st \pm
0.041 \sy \] where the uncertainties are dominated by the systematic
errors.

The Coulomb repulsive interaction opposes the Bose-Einstein
enhancement in the low $Q_3$ region and therefore it is reasonable
that in our analysis the Coulomb correction increased the $\lambda_3$
value. This increase amounts to about $9\%$. On the other hand, the
Coulomb correction has a smaller effect on the $r_3$ value which is
lowered by about $6\%$. Accounting for the three-pion purity the BEC
strength amounts to $\lambda_3^{\rm pure} = 0.707 \pm 0.014 \st \pm
0.078 \sy$ for a $100\%$ pure $\pipipi$ system.

A relation between the two-pion and the three-pion emitter dimensions
was discussed in reference \cite{juricic}. We tested this relation by
using the present result and that obtained in the latest OPAL two-pion
BEC analysis \cite{newm} where approximately the same data sample was
used. The proposed relation between $r_2$ and $r_3$, expressed in
Eq.~\ref{eq_r3e}, is in good agreement with our results.

\section*{Acknowledgements}

We particularly wish to thank the SL Division for the efficient
operation of the LEP accelerator at all energies and for their
continuing close cooperation with our experimental group. We thank
our colleagues from CEA, DAPNIA/SPP, CE-Saclay for their efforts over
the years on the time-of-flight and trigger systems which we continue
to use. In addition to the support staff at our own
institutions we are pleased to acknowledge the\\
Department of Energy, USA,\\
National Science Foundation, USA,\\
Particle Physics and Astronomy Research Council, UK,\\
Natural Sciences and Engineering Research Council, Canada,\\
Israel Science Foundation, administered by the Israel
Academy of Science and Humanities,\\
Minerva Gesellschaft,\\
Benoziyo Center for High Energy Physics,\\
Japanese Ministry of Education, Science and Culture (the Monbusho) and
a grant under the Monbusho International
Science Research Program,\\
German Israeli Bi-national Science Foundation (GIF),\\
Bundesministerium f\"ur Bildung, Wissenschaft,
Forschung und Technologie, Germany,\\
National Research Council of Canada,\\
Research Corporation, USA,\\
Hungarian Foundation for Scientific Research, OTKA T-016660,
T023793 and OTKA F-023259.\\

\end{document}